\documentclass[twocolumn]{aastex701}

\usepackage{graphicx}
\usepackage{xcolor}

\newcommand{\ath}{{\sc athena++} }
\newcommand{\mesa}{{\sc mesa} }

\begin{document}
\title{From Inspiral to Outburst: Modeling a Potential Luminous Red Nova from Betelgeuse}

\author[orcid=0000-0001-5782-3719,sname='Naunet Leonhardes-Barboza']{Naunet Leonhardes-Barboza}
\affiliation{Department of Astronomy and Astrophysics, University of California, Santa Cruz, CA 95064, USA}
\email[show]{njleonha@ucsc.edu}  

\author[orcid=0000-0003-3062-4773, sname='Andrea Antoni']{Andrea Antoni} 
\affiliation{Center for Computational Astrophysics, Flatiron Institute, New York, NY 10010, USA}
\email[]{aantoni@flatironinstitute.org}

\author[orcid=0000-0002-1417-8024, sname='Morgan MacLeod']{Morgan MacLeod} 
\affiliation{Center for Astrophysics, Harvard \& Smithsonian, 60 Garden Street, Cambridge, MA 02138, USA}
\affiliation{Institute for Theory and Computation, Harvard \& Smithsonian, 60 Garden Street, MS-51, Cambridge, MA 02138, USA}
\email[]{morgan.macleod@cfa.harvard.edu}

\author[orcid=0000-0000-0000-0000, sname='Angela Twum']{Angela Twum} 
\affiliation{Department of Astronomy and Astrophysics, University of California, Santa Cruz, CA 95064, USA}
\email[]{atwum@ucsc.edu}

\author[orcid=0000-0003-2558-3102, sname='Enrico Ramirez-Ruiz']{Enrico Ramirez-Ruiz} 
\affiliation{Department of Astronomy and Astrophysics, University of California, Santa Cruz, CA 95064, USA}
\email[]{enrico@ucolick.org}

\begin{abstract}
Recent evidence suggests that Betelgeuse may host a stellar companion whose orbit will decay through tidal interactions, ultimately leading to engulfment and merger within the next several thousand years. If confirmed, this system could represent the nearest and best-characterized progenitor of a future luminous red nova (LRN). We investigate whether the shock-breakout and cooling emission preceding the LRN can be used to constrain the properties of the companion and its disruption within the stellar envelope. Using binary stellar evolution calculations, we evolve Betelgeuse with companions of $0.7$, $1.0$, and $2.0\,M_\odot$ to the onset of common envelope evolution (CEE). We then perform one-dimensional radiation-hydrodynamic simulations that follow companion disruption, energy deposition, and shock propagation through the envelope. We find that the peak luminosity increases systematically with companion mass, from $2.44\times10^{41}$ to $6.17\times10^{41}\ {\rm erg\;s^{-1}}$ across the $0.7$--$2.0 \; M_\odot$ models, while the duration increases more modestly from $1.48$ to $1.99$ days and attains apparent magnitudes brighter than Venus as viewed from Earth. Tidal disruption contributes both kinetic energy and mass to the envelope, linking the companion's structure and mass to the energetics of the shock-heated ejecta. The monotonic luminosity sequence demonstrates that the early breakout emission can retain direct information about the engulfed companion. The interaction ejects $6.15$--$11.19\,M_\odot$ of hydrogen-rich material, substantially modifying both the envelope structure and circumstellar environment prior to core collapse.
\end{abstract}



\section{Introduction} 

Betelgeuse ($\alpha$ Orionis), the brilliant red supergiant marking Orion's shoulder, is among the most intensively examined stars in the sky. Situated approximately $172 \pm 13$ pc away \citep{MacLeod2025}, Betelgeuse offers a unique view into the late evolutionary stages of massive stars, providing invaluable insight into pre-supernova processes. Thought to have originated from a zero-age main sequence (ZAMS) progenitor of $17.5\pm 2M_\odot$ \citep{MacLeod2025}, Betelgeuse now possesses a radius of $782\pm 55,R_\odot$ \citep{MacLeod2025}, an effective temperature of $\approx 3600 \pm 25$ K \citep{Levesque2020,Harper2020}, and a luminosity near $10^5 \, L_\odot$ \citep{MacLeod2025,Goldberg2024,Levesque2020}.

Betelgeuse has held a prominent place not only in the development of stellar astrophysics but also in the tapestry of human culture and history. This name derives from the Arabic Yad al-Jawzā’, meaning ``the hand of the central one,'' referencing Orion’s celestial placement \citep{Allen1899}. The Persian astronomer al-Sufi cataloged the star during the Islamic Golden Age in his {\it Book of the Fixed Stars} (ca. 964 CE), which combined the Ptolemaic stellar catalog with indigenous Arabic star names and became an influential astronomical reference throughout the Islamic world and later Europe. The bright star's color evolution has been followed extensively over two millennia, with recorded observations in Greece, Rome, and China throughout history \citep{Neuhauser2022}.

Betelgeuse features prominently in Indigenous Knowledge Systems. In Hawaiian star knowledge, `Oihana Kilo H\={o}k\={u}, Betelgeuse is known as Kauluaakoko. It forms a part of the constellation Ka Heihei o N\={a} Keiki (Orion), which belongs to the traditional Polynesian wayfinding and seasonal reckoning system \citep{Makemson1941}. In Māori astronomy, the star is recorded as Pūtara \citep{Harris2013}. In the astronomical traditions of the Havasupai of the southwestern United States, Betelgeuse was identified with Coyote Man within their interpretation of Orion, whose seasonal appearance helped mark the annual calendar \citep{Whiting1985}. Among the Lacand\'on Maya, the star was known as ch\"ak tulix, the ``red dragonfly,'' explicitly identifying the star by its conspicuous color \citep{Milbrath1999}. Betelgeuse also forms part of the region of Orion associated in Maya traditions with the celestial turtle, creation imagery, and seasonal cycles of maize cultivation \citep{Milbrath1999}. The variability of this red supergiant is noted in Indigenous Australian oral stories as well \citep{Hamacher2018, Schaefer2018}. The many understandings of this red star\footnote{In this paper, we primarily refer to this star as Betelgeuse.} illustrate not only its longstanding importance to astronomical observation, but also to navigation, seasonal knowledge, and cultural connection to the night sky.

Betelgeuse serves as a crucial laboratory for investigating stellar structure, convective dynamics, and mass-loss mechanisms in massive stars (see \citealt{Wheeler2023} and \citealt{Dupree2025} for recent reviews). Its semi-regular variability in brightness, also recorded in the 19th century by Sir John Herschel, is now attributed to large-scale pulsations and dynamic convective activity \citep{Herschel1840}. High-resolution interferometry and spectroscopy with instruments such as the VLTI, ALMA, and the Hubble Space Telescope have resolved Betelgeuse’s extended photosphere, revealing asymmetric surface features, giant convective cells, and a complex circumstellar medium (CSM) \citep{Gilliland1996,Lobel2000, Lobel2001, Montarges2016,Kervella2018}. For decades, researchers debated whether Betelgeuse hosts a binary companion, motivated by unusual radial velocity variations, long secondary photometric periods, and its unexpectedly rapid surface rotation. Recent independent analyses by \citet{Goldberg2024} and \citet{MacLeod2025} provided compelling evidence that these phenomena can be explained by a low-mass stellar companion orbiting within Betelgeuse’s extended chromosphere. This companion is now designated as $\alpha$ Orionis B, or Siwarha, meaning ``her bracelet'' in Arabic. Both models favor a companion in a $\approx 2170$ day ($\approx 6$ yr) orbit at a separation of $\approx 8.6$ AU with predicted masses of respectively $M_c \sin i = 1.17 \pm 0.07 \, M_\odot$ and $M_c = 0.60 \pm 0.14 \pm 0.07 \, M_\odot$ with an inclination $i = 98^{\circ}$, where tidal interactions naturally spin up the envelope while driving orbital decay on timescales of order $\approx 10^4$ yr. Ultraviolet and X-ray observations further disfavor a compact object companion, instead supporting the presence of a low-mass stellar companion \citep{OGrady2025}.

This binary interpretation gained further support with the probable direct-imaging detection reported by \citet{Howell2025}. Using the ‘Alopeke speckle imager on Gemini North, they identified a faint source at a separation of $\approx 52$ milliarcseconds from Betelgeuse at $1.5\sigma$, closely matching the predicted orbital position and brightness of the proposed companion, and estimated masses of $\approx 1.4 \;\rm{ to} \; 2 M_\odot$ using MIST \citep{Choi2016, Dotter2016}. Direct imaging by \citet{Montarges2026} using the Very Large Telescope SPHERE-ZIMPOL at a similar separation at $6.1\sigma$ yielded companion masses of $\approx 2.6 \;\rm{ to} \; 3.1 M_\odot$. Although these detections remain tentative and the companion masses are model-dependent, the agreement between the observed position and prior dynamical predictions substantially strengthens the case for an interacting binary system approaching CEE.

The predicted engulfment of Betelgeuse’s companion places the system within the broader class of luminous red novae (LRNe), transients powered by unstable mass transfer, CEE, and stellar mergers. Observed LRNe such as V838 Mon and V1309 Sco have revealed a diverse population of merger-powered eruptions spanning a wide range of luminosities, durations, and progenitor properties \citep{Munari2002,2011A&A...528A.114T,Twum2026}. However, most known systems are either poorly constrained prior to eruption or originate from comparatively compact progenitors. Betelgeuse likely represents one of the most evolved progenitors in the LRNe landscape currently constrained by pre-outburst observations \citep{Twum2026}. 

Betelgeuse’s ultimate fate is expected to be a hydrogen-rich (Type II) core-collapse supernova, likely within the next $\approx 10^5$ years, although the precise timing remains uncertain and model dependent \citep{Meynet2013}. Before core collapse, however, the inspiral and engulfment of its companion may substantially reshape the CSM, alter the star’s final evolutionary pathway, and produce a luminous pre-supernova transient visible from Earth. Such an event would offer an unprecedented opportunity to directly connect CEE, stellar mergers, and LRN with the late stages of massive-star evolution.

By evolving Betelgeuse to the onset of engulfment and merger, we estimate the properties of the resulting transient, with particular emphasis on the shock-powered blue peak produced by rapidly expanding ejecta during the earliest stages of common-envelope interaction. A central question is whether this early emission retains observable information about the companion itself. Because the Betelgeuse donor structure at engulfment remains nearly unchanged across the companion masses considered here, our calculations provide a controlled experiment for determining how companion mass, tidal disruption, and energy deposition are mapped into the luminosity and duration of the emerging transient.

We combine binary stellar evolution calculations with one-dimensional radiation hydrodynamics simulations that follow the inspiral-driven energy release, shock propagation through the envelope, and shock breakout at the stellar surface. These calculations provide one of the first predictive models for a nearby LRN arising from a massive red supergiant progenitor. Section \ref{sec:Sec1} describes the pre-merger binary stellar evolution, Section \ref{sec:merger} reviews the hydrodynamical setup, Section \ref{sec:Results} presents the resulting transient properties, and Section \ref{sec:Discussion} discusses the results and implications for LRNe and Betelgeuse’s future evolution.

\begin{figure*}
\centering
\includegraphics[width=7.1in]{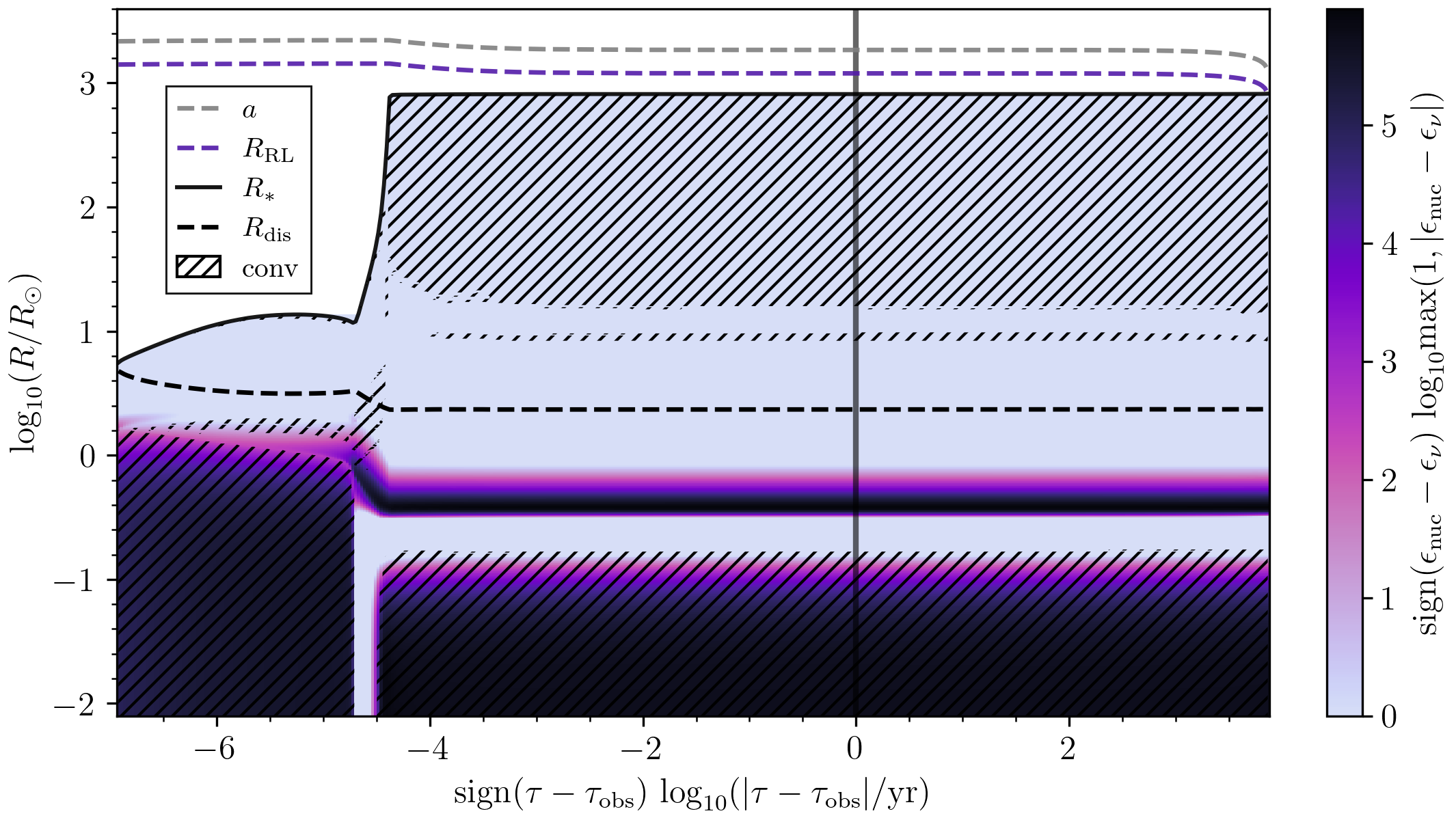}
\caption{
Kippenhahn-like diagram illustrating the late-stage evolution of Betelgeuse and a $0.7M_\odot$ companion in \mesa. The diagonal hatching denotes convective mixing regions. The color scale traces regions of nuclear burning, shown as the logarithmic difference between local nuclear energy generation and neutrino cooling. The x-axis gives the logarithmic time relative to the calibrated present-day model, marked by the vertical black line at $0$. Positive values correspond to future evolution toward engulfment, while negative values trace the preceding evolution. As Betelgeuse expands during core helium burning, tidal interactions drive orbital decay and reduce its Roche-lobe radius, $R_{\rm RL}$, toward its stellar surface, $R_*$. The tidal disruption radius, $R_{\rm dis}$, remains concentrated near the stellar core, indicating that the companion survives until deep inspiral before being disrupted.}
\label{fig:onset_RLOF}
\end{figure*}
\section{Evolution Toward Engulfment \label{sec:Sec1}}
Before modeling the hydrodynamic inspiral and merger, we first evolve the Betelgeuse binary system toward the onset of engulfment using one-dimensional stellar evolution calculations. Our primary goal is to determine the structure of Betelgeuse at the onset of CEE and to construct physically motivated initial conditions for the subsequent radiation hydrodynamics simulations. 

We build upon the co-evolutionary binary framework presented by \citet{MacLeod2025}, who modeled Betelgeuse and its proposed low-mass companion using using the Modules for Experiments in Stellar Astrophysics ({\tt MESA}) version r24081 \citep{2011ApJS..192....3P,2013ApJS..208....4P,2015ApJS..220...15P,2018ApJS..234...34P,2019ApJS..243...10P}. Their models were calibrated to reproduce the presently observed properties of the Betelgeuse system, including the stellar luminosity, effective temperature, spin period of $\approx 36$ yr, the observed long secondary period interpreted as the orbital period of the companion ($\approx 2110$ days), and the fundamental pulsation mode period of $\approx 410$ days.

We evolve Betelgeuse with $M_{\rm ZAMS}=18.8M_\odot$ and solar metallicity ($Z=0.02$) together with a low-mass companion in a circular orbit. We use the \texttt{approx21\_cr60\_plus\_co56.net} reaction network to follow massive stellar evolution through the late stages of core helium burning. Similar to \citet{Antoni2025}, we adopt the OPAL \citet{Asplund2009} opacity tables (\texttt{kap\_file\_prefix = oplib\_agss09}), which are most consistent with the opacity treatment adopted in the hydrodynamic calculations described in Section \ref{sec:merger}. We initialize the system with an orbital period of $2650$ days and adopt \texttt{mixing\_length\_alpha = 2.1}, following \citet{MacLeod2025} and \citet{Joyce2020}. We do not include overshooting in this work because its treatment remains uncertain and can significantly alter convective damping efficiencies during tidal evolution \citep{Brott2011,Martins2013,Aerts2019}.

The future orbital evolution of the Betelgeuse system is governed by tidal interactions between the companion and the extended convective envelope of the red supergiant. As in \citet{MacLeod2025}, we adopt the equilibrium-tide formalism with convective damping as implemented in \mesa following \citet{2002MNRAS.329..897H}. We set the \texttt{tidal\_reduction = 2} to follow \citet{Goldreich1977}'s framework of coupling tides to convection using mixing length theory. In this framework, tidal torques exchange angular momentum between the orbit and the stellar envelope, allowing synchronization and orbital circularization as Betelgeuse evolves across the red-supergiant phase. We treat the rotation of Betelgeuse as a solid body throughout its evolution.
Starting from the calibrated present-day configuration, we evolve the system forward in time to the onset of Roche-lobe overflow (RLOF), which we adopt as the onset of engulfment. At this stage, the mass of Betelgeuse is $M_* = 18.25 M_\odot$ and the radius is $R_* = 819.74 R_\odot$. To explore the sensitivity of the future evolution to the uncertain companion properties and tidal evolution, we vary the companion mass between $0.7M_\odot$, $1.0M_\odot$, and $2.0M_\odot$. These models produce times to engulfment from the present epoch of $7262$, $5702$, and $4293$ yr, respectively. We note that because the coupling of tidal flows to convection is uncertain at the order of magnitude (or more) level \citep{2014ARA&A..52..171O}, these model-predicted times are not intended to be precise reflections of the evolution of the realistic system. 

Despite the substantial variation in inspiral timescale, the internal structure of Betelgeuse at engulfment remains comparatively insensitive to the companion mass and orbital evolution history. In all cases, Betelgeuse enters its core helium-burning phase prior to engulfment and develops a deeply convective, weakly bound red-supergiant envelope. For the remainder of this work, we use the co-evolved donor profiles at engulfment to initialize the hydrodynamic calculations. Because the Betelgeuse structure remains rather similar across the explored companion masses i.e. staying in the He-burning phase which lasts $\gtrsim 10^4$ years,  the choice of companion mass primarily affects the time to engulfment ($\approx10^3$ years) rather than the donor profile itself. This similarity allows differences among the subsequent shock-breakout calculations to be attributed primarily to the companion properties and their associated energy and mass deposition, rather than to substantial differences in the donor structure.

Figure \ref{fig:onset_RLOF} presents the late-stage evolution of the Betelgeuse system in a Kippenhahn-like diagram. As Betelgeuse expands during core helium burning, tidal interactions increasingly influence the orbital evolution, leading to orbital decay and a reduction of the Roche-lobe radius toward the stellar surface. The tidal disruption radius remains concentrated near the stellar core throughout the evolution, indicating that the companion survives intact until deep inspiral. During this phase, Betelgeuse loses only a small fraction of its total mass through winds.

The efficiency of tidal dissipation in convective stellar envelopes remains highly uncertain theoretically. In equilibrium-tide models, dissipation depends on the interaction between turbulent convection and large-scale tidal flows \citep{2014ARA&A..52..171O}, therefore different prescriptions for turbulent viscosity and convective damping can alter inspiral timescales substantially \citep{2020MNRAS.498.2270B}. It is useful to compare the tidal dissipation power to the stellar luminosity of Betelgeuse observed today ($L_* \approx 3.8-4 \times 10^{38} \,\rm{erg \, s^{-1}}$). The total orbital energy at a separation of $a \approx R_*$ is $E_{\rm orb} \approx G M_1 M_2 / a \sim  G M_1 M_2 / R_* \approx 8.5 \times 10^{46} (M_2 \,  / \, M_\odot) \, \rm{erg}$. Following \citet{MacLeod2026} and \citet{Millholland2025} for the fiducial $M_2 = 1.0M_\odot$ case, we obtain a strong tidal quality factor of $Q'_* \approx 8 \times 10^3$ and an orbital decay timescale of $\tau_{\rm decay} \approx 10^4$ years. This leads to a luminosity due to tides of $L_{\rm tide} \sim E_{\rm orb}\,/\, \tau_{\rm decay} \approx 10^{35} \,\rm{erg \, s^{-1}} \lesssim L_*$. While the orbit decays on a relatively fast timescale, the intrinsic stellar output is high enough we do not expect Betelgeuse to become brighter because of this extra power.

As Betelgeuse expands and tidal torques increasingly couple the orbit to the extended convective envelope, the system approaches the regime where synchronous rotation can no longer be maintained. Binary systems containing low-mass companions embedded within massive extended envelopes are susceptible to the Darwin instability \citep{Darwin1879,MacLeod2017}, which occurs when the orbital angular momentum becomes insufficient to sustain co-rotation of the envelope. Once this threshold is crossed, angular momentum transfer from the orbit to the envelope drives rapid orbital decay, producing an abrupt transition from quasi-secular tidal evolution to dynamical inspiral and CEE. We therefore interpret the onset of Roche-lobe overflow as the beginning of this runaway engulfment phase and terminate the stellar evolution calculations at this point.

In the following section, we use the co-evolved donor profiles described here as the initial conditions for radiation hydrodynamics simulations designed to follow the inspiral-powered energy deposition, shock propagation, tidal-disruption energy release, and emergent shock-breakout signature of the engulfment event.

\section{The Merger} \label{sec:merger}

Following the onset of engulfment, the companion first slowly, then increasingly rapidly spirals through Betelgeuse's extended envelope, depositing orbital energy into the surrounding gas before ultimately being tidally disrupted near the stellar core. The resulting debris thermalizes within the envelope, converting orbital kinetic energy into heat and driving a pressure wave that steepens into an outward-propagating shock. To follow the subsequent evolution of the envelope and predict the emergent shock-breakout transient, we perform one-dimensional radiation hydrodynamics simulations using the Eulerian code \ath \citep{Stone2020}.

The simulations solve the coupled equations of inviscid hydrodynamics and frequency-integrated radiation transport \citep[e.g., ][]{Jiang2014,Jiang2021,Antoni2025} assuming spherical symmetry. For the gas, we adopt a $\gamma$-law equation of state using a $\gamma = 5/3$. The gamma-law EOS is sufficient until H-recombination starts to become important for the light curve, but we stop the models before then. We use the LANL atomic opacities \citep{Colgan2016}, and at low temperatures, these are smoothly connected to the Rosseland- and Planck-mean opacity tables of \citet{Ferguson2005}. The density, pressure, and radiative luminosity profiles from the \mesa donor model at the onset of engulfment are mapped onto the \ath\ grid, and an isothermal atmosphere with temperature $T=T_{*}$ is appended at the stellar surface.

The computational domain extends from $r_{\rm in}=10^{-1}R_\odot$ to $r_{\rm out}=10^{5}R_\odot$. We excise the innermost region of the star and apply outflow boundary conditions at both boundaries. Because our primary interest is the shock breakout and early transient emission, the nuclear burning luminosity generated interior to $r_{\rm in}$ is not included in the calculation. We also neglect convective transport in the \ath calculation because the injected energy drives a supersonic, dynamical response for which convective transport is slow compared with the shock propagation timescale.

\begin{figure*}[ht!]
\centering
\includegraphics[width=7.1in]
    {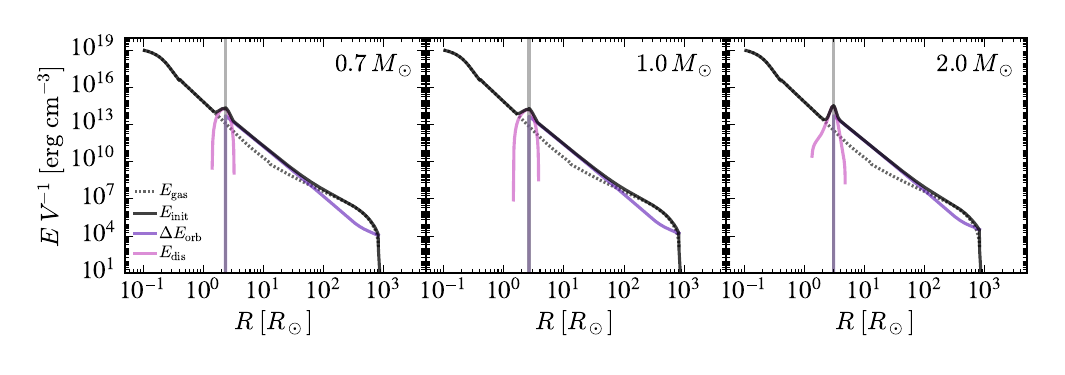}
    \caption{ Energy density profiles used to initialize the merger simulations. The dotted curve shows the gas energy density ($E_{\rm gas}$) of the donor at the onset of engulfment obtained from our \mesa models. The purple solid curve  shows the orbital energy released during inspiral ($\Delta E_{\rm orb}$), while the pink solid curve  shows the kinetic energy associated with the tidally disrupted companion debris ($E_{\rm dis}$). The solid black curve represents the total injected energy density deposited into the stellar envelope ($E_{\rm init} = \Delta E_{\rm orb}+E_{\rm dis} + E_{\rm{gas}}$). Vertical lines mark the tidal disruption radius, $r_{\rm dis}$, for each companion mass. The figure demonstrates that both the orbital and disruption energy are deposited deep within the stellar interior, well below the extended convective envelope, ensuring that the subsequent evolution proceeds through shock propagation rather than prompt radiative escape. The outer isothermal atmosphere appended to the \ath\ models is omitted for visual clarity.}
    \label{fig:energy_deposition}
\end{figure*}


As the companion spirals inward through the envelope, orbital energy is transferred to the surrounding gas. During the final stages of inspiral, the orbital decay timescale becomes short compared to the thermal adjustment time of the envelope, and the released energy is expected to drive a strong pressure wave that subsequently propagates through the star. Because our primary interest is the resulting shock breakout and early transient emission, we approximate the inspiral as an impulsive energy deposition event. This approximation is motivated by the separation of timescales between energy deposition near the disruption radius, $r_\mathrm{dis}$ (in Equation \ref{eq:rdis}), and the global dynamical response of the envelope: \(t_{\rm dep}\approx P_{\rm orb}(r_{\rm dis})\ll t_{\rm dyn}(R_\ast)\). The envelope therefore cannot respond globally during the final phase of energy deposition. The deposited orbital energy is computed from the radial change in the companion's orbital binding energy,
\begin{equation}
\frac{dE_{\rm orb}}{dV}
=
-\frac{Gm_c}{2}
\left[
\frac{\rho(r)}{r}
-
\frac{M_{\rm enc}(r)}{4\pi r^4}
\right],
\end{equation}
where $dE_{\rm orb}/dV$ is the orbital energy released per unit volume, $m_c$ is the companion mass, $M_{\rm enc}(r)$ is the enclosed mass of the donor, and $\rho(r)$ is the local density profile. This expression follows from differentiating the orbital binding energy, $E_{\rm orb}=-Gm_cM_{\rm enc}(r)/(2r)$, with respect to radius and depositing the released energy locally within the stellar envelope.

As the inspiral proceeds, the companion eventually reaches the tidal disruption radius, $r_{\rm dis}$, defined by
\begin{equation}
\label{eq:rdis}
\rho_{\rm enc}(r_{\rm dis})
=
\frac{\rho_c}{2},
\end{equation}
where $\rho_{\rm enc}$ is the mean enclosed density of the donor and $\rho_c$ is the average density of the companion. Interior to $r_{\rm dis}$, the companion is assumed to be tidally disrupted and no longer behaves as a coherent gravitating body. The disrupted material retains substantial kinetic energy associated with its orbital motion at the point of disruption. As the debris interacts with the surrounding envelope, this kinetic energy is dissipated and thermalized, depositing additional energy and momentum into the star. Because the disruption process is intrinsically multidimensional and occurs deep within an optically thick environment, we approximate its effect in our one-dimensional framework by depositing the mass and kinetic energy of the disrupted companion locally around the tidal disruption radius.

To model this process, we assume that each disrupted mass element retains the local orbital velocity at the tidal disruption radius. The kinetic energy density associated with the debris is therefore
\begin{equation}
\frac{dE_{\rm dis}}{dV}
=
\frac{1}{2}\rho_{\rm dis}(r)\,v_{\rm orb}^2(r),
\end{equation}
where $v_{\rm orb}(r)=\sqrt{GM_{\rm enc}(r)/r}$ is the local orbital velocity and $\rho_{\rm dis}$ describes the radial distribution of disrupted companion material. The total deposited energy is then
$ E_{\rm total}=\Delta E_{\rm orb}
+ E_{\rm dis}.$
The disrupted companion is projected onto the one-dimensional grid in a manner that conserves both mass and kinetic energy, yielding the radial debris density profile $\rho_{\rm dis}(r)$. The disruption energy is not an independent energy source. Rather, it represents the kinetic energy carried by companion material at the moment of tidal disruption and therefore captures the final stages of orbital energy release that are not included once the inspiral calculation is terminated at $r_{\rm dis}$.

We perform three simulations with companion masses of $0.7M_\odot$, $1.0M_\odot$, and $2.0M_\odot$. Figure~\ref{fig:energy_deposition} shows the corresponding energy deposition profiles. The shapes of the disruption kernels reflect the internal structures of the companions. The $0.7M_\odot$ and $1.0M_\odot$ companions remain in their pre-main-sequence phase at the time of engulfment and therefore possess extended, low-density envelopes, while the $2.0M_\odot$ companion has already reached the main sequence and exhibits a more centrally concentrated structure.
\begin{figure*}
    \centering
    \includegraphics[width=7.0in]{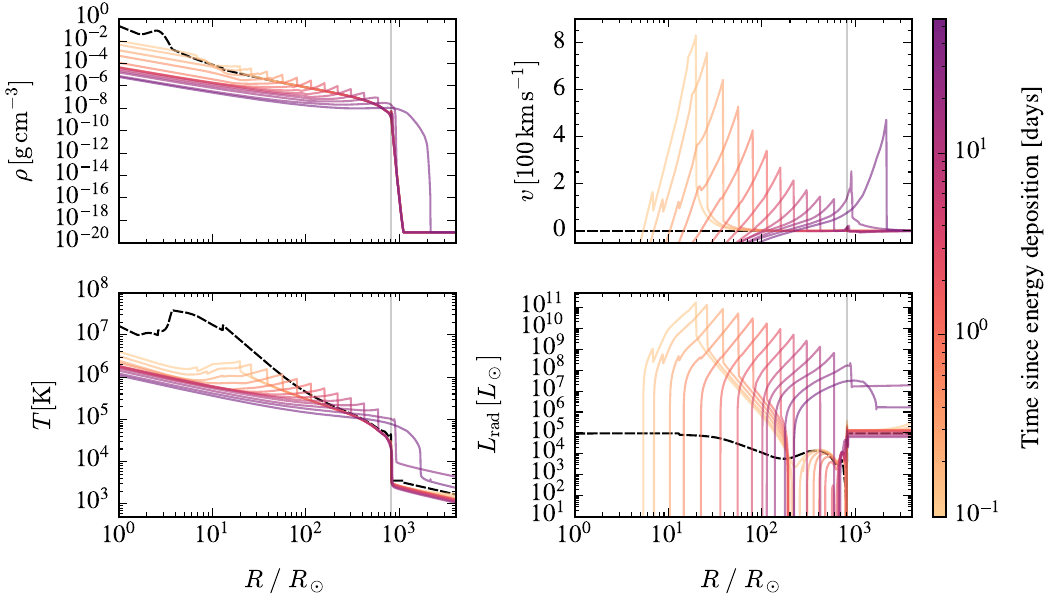}
    \caption{Evolution of the fiducial $1.0M_\odot$ companion model following the deposition of $E_{\rm total}=8.05\times10^{48}$ erg. The dashed curves show the initial conditions immediately after energy deposition, while the colored curves represent logarithmically spaced snapshots in time. From top to bottom, the panels show the density, velocity, radiative luminosity ($L_{\rm rad}$), and gas temperature profiles. The vertical black line marks the stellar surface of the mapped \mesa progenitor model. The deposited energy launches a shock that propagates through the extended red-supergiant envelope, accelerates through the steep outer density gradient, and ultimately breaks out at the stellar surface.
}
    \label{fig:time_evolve}
\end{figure*}
In order of increasing companion mass, the companion radii are $0.93R_\odot$, $1.18R_\odot$, and $1.723R_\odot$, yielding tidal disruption radii of $2.33R_\odot$, $2.62R_\odot$, and $3.05 R_\odot$, respectively\footnote{We use 99.9 \% of the companion mass to calculate the tidal radii.}. The corresponding orbital energy release for each companion is $\Delta E_{\rm orb}=3.09\times10^{48}$, $3.95\times10^{48}$, and $6.85\times10^{48}$ erg, while the kinetic energy of the disrupted debris is $E_{\rm dis}=3.20\times10^{48}$, $4.10\times10^{48}$, and $7.02\times10^{48}$ erg. The total deposited energies are therefore $E_{\rm total}=6.29\times10^{48}$, $8.05\times10^{48}$, and $1.39\times10^{49}$ erg. Despite being disrupted at a larger radius, the $2.0 \, M_\odot$ companion supplies substantially more total energy because of its larger mass. The total deposited energy therefore increases monotonically with companion mass across the three models. As we show in Section 4, this energetic sequence is directly reflected in the peak breakout luminosity.

For all three models, the deposited energy is concentrated deep within the stellar interior, interior to the base of the convective hydrogen envelope located at $r_{\rm H, conv} = 12.93R_\odot$ with a total mass $M_{\rm conv}=12.45M_\odot$. The deposition region therefore lies well beneath the stellar photosphere and at optical depths sufficiently large that the released energy cannot escape directly as radiation. Instead, the rapid deposition of several $10^{48}$ erg drives a strong pressure wave that steepens into a shock as it propagates outward through the extended red-supergiant envelope. Owing to Betelgeuse's enormous radius and weakly bound hydrogen envelope, the shock traverses nearly three orders of magnitude in radius before reaching the stellar surface, where its emergence produces a luminous shock-breakout transient. The propagation and breakout of this shock are the primary focus of the following section.

\section{From Engulfment to Shock Breakout\label{sec:Results}}
 
The deposition of orbital and disruption energy deep within Betelgeuse's envelope drives a strong pressure wave that rapidly steepens into a shock. Because the energy is injected at radii of only a few $R_\odot$, far below the stellar photosphere, the subsequent evolution is governed by the propagation of this shock through the extended red-supergiant envelope. In this section, we first examine the hydrodynamic evolution of the shock and then discuss the resulting shock-breakout lightcurves.

Figure~\ref{fig:time_evolve} shows the evolution of the fiducial $1.0M_\odot$ model following the deposition of orbital and disruption energy. The injected energy launches a strong shock that propagates outward through the envelope. As the shock moves into the steep density gradient of the outer envelope, it accelerates and heats the surrounding gas, producing a radiation-dominated shock front that carries energy toward the stellar surface. After approximately 28 days, the shock reaches the stellar surface and photons begin to escape. This transition is visible in the final snapshots, where the luminosity becomes approximately constant with radius outside $\approx 10^3R_\odot$, indicating freely streaming radiation.

The duration of the shock-breakout and cooling emission provides a useful diagnostic of the physical processes governing the release of energy from the shock-heated envelope. For extended red-supergiant progenitors such as Betelgeuse, the observable duration may be compared to three characteristic timescales: the progenitor light-crossing time, $t_{\rm lc}$, the diffusion time through the breakout layer, $t_{\rm diff}$, and the expansion time of the ejecta, $t_{\rm exp}$ \citep{Piro2013}. For Betelgeuse, the light-crossing time is $t_{\rm lc}=R_*/c\approx 40$ min. Using Equations (6) and (8) of \citet{Piro2013}, we estimate $t_{\rm diff}=0.86$ d and $t_{\rm exp}=10.4$ d for our fiducial $1.0M_\odot$ model, yielding the ordering $t_{\rm lc}\ll t_{\rm diff}<t_{\rm exp}$. Because $t_{\rm lc}\ll t_{\rm diff}$, geometric light-travel-time effects make a negligible contribution to the observed duration of the blue peak. Furthermore, the fact that $t_{\rm exp}$ substantially exceeds $t_{\rm diff}$ indicates that adiabatic expansion should not strongly suppress the emerging emission. As we show below, the duration of the blue peak measured in the radiation-hydrodynamic simulations is comparable to the analytic diffusion timescale, consistent with the breakout and cooling emission being regulated primarily by radiative diffusion through the shock-heated envelope.




\begin{figure}
\includegraphics[width=3.35in]{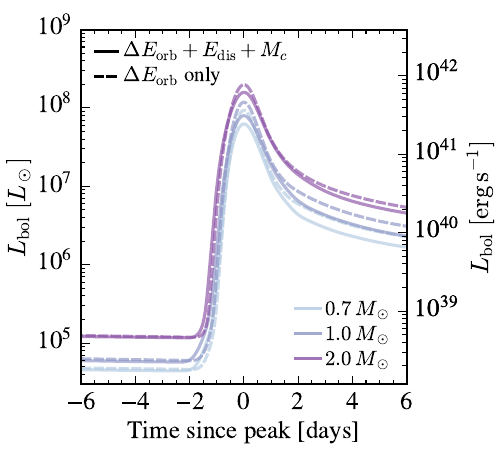}    
\caption{Bolometric luminosity as a function of time for the three companion-mass models. The light curves are aligned such that the luminosity peak occurs at $t=0$. Solid curves correspond to the fiducial models, which include the companion mass together with both the orbital energy released during inspiral and the kinetic energy deposited by the tidally disrupted debris, $\Delta E_{\rm orb} + E_{\rm dis} + M_c$. Dashed curves show models in which only the inspiral energy, $\Delta E_{\rm orb}$, is deposited. Comparing the two cases illustrates how the additional mass and energy associated with tidal disruption modify the specific energy of the ejecta and consequently the luminosity and duration of the breakout and cooling emission.
}
\label{fig:L_peak}
\end{figure}

Figure~\ref{fig:L_peak} shows the bolometric light curves of the three companion-mass models, aligned such that the maximum luminosity occurs at $t=0$. The solid curves correspond to our fiducial models, which include both the orbital energy released during inspiral and the mass and kinetic energy of the tidally disrupted companion. The dashed curves show models in which only the inspiral energy, $\Delta E_{\rm orb}$, is deposited. This comparison isolates the impact of tidal disruption on the resulting transient.

For the fiducial models, we find peak luminosities of $2.44\times10^{41}$, $3.11\times10^{41}$, and $6.17 \times 10^{41}\; {\rm erg\ s^{-1}}$ for the $0.7M_\odot$, $1.0M_\odot$, and {$2.0M_\odot$} companions, respectively. The peak luminosity therefore increases monotonically with companion mass, rising by a factor of \(\simeq2.5\) between the \(0.7\) and \(2.0\,M_\odot\) models. This sequence closely follows the increase in total deposited energy and demonstrates that the breakout luminosity retains information about the engulfed companion.

We characterize the duration of the shock breakout and cooling emission using the full width at half maximum (FWHM) of the blue peak, obtaining durations of 1.48, 1.68, and $1.99$ days. By contrast, the FWHM varies by only $\sim 35 \%$, suggesting that peak luminosity is more sensitive to companion mass than the characteristic duration of the breakout.

Assuming that the dynamic photosphere radiates approximately as a blackbody, the corresponding peak magnitudes are $M_V=-13.47$, $-13.61$, and $-13.97$ in Bessell $V$, and $M_r=-13.32$, $-13.45$, and $-13.80$ in LSST-$r$. In all cases, the transient briefly exceeds Betelgeuse's present luminosity by several orders of magnitude. The merger also ejects a substantial fraction of the hydrogen-rich envelope. Using the criterion $E_{\rm kin}+E_{\rm grav}>0$, we find unbound ejecta masses of 6.15, 8.11, and $11.19M_\odot$ for the $0.7M_\odot$, $1.0M_\odot$, and $2.0 M_\odot$ models, respectively. The corresponding asymptotic mass-weighted ejecta velocities are 194, 207, and $246 \; {\rm km \; s^{-1}}$. These values are broadly consistent with the outflow velocities inferred for LRNe \citep[e.g.,][]{Munari2002,2011A&A...528A.114T,Twum2026} and imply the removal of a significant fraction of Betelgeuse's hydrogen envelope. A more detailed discussion of the ejecta energetics and comparison with analytical common-envelope prescriptions is presented in the Appendix.

The comparison between the solid and dashed curves further illustrates the importance of tidal disruption in lowering the peak luminosity. Disruption injects both additional kinetic energy and the companion mass itself into the shocked material. Although this mass loading reduces the specific energy relative to adding the same energy without additional mass, the increase in total deposited energy with companion mass dominates the peak luminosity across the models considered here. The peak luminosity scales with companion mass as $L_{\rm peak} \propto M_{\rm comp}^{0.93}$ when both inspiral and tidal disruption are included, compared to $L_{\rm peak} \propto M_{\rm comp}^{0.73}$  when disruption energy and companion mass deposition are neglected. Both dependencies are steeper than the $L_{\rm peak} \propto M_{\rm comp}^{2/3}$ scaling expected from orbital energy alone. The stronger dependence in the fiducial models highlights the additional role of tidal disruption and mass deposition in determining the energetics and luminosity of the emerging transient.

\begin{figure*}
    \centering
    \includegraphics[width=7.1in]
    {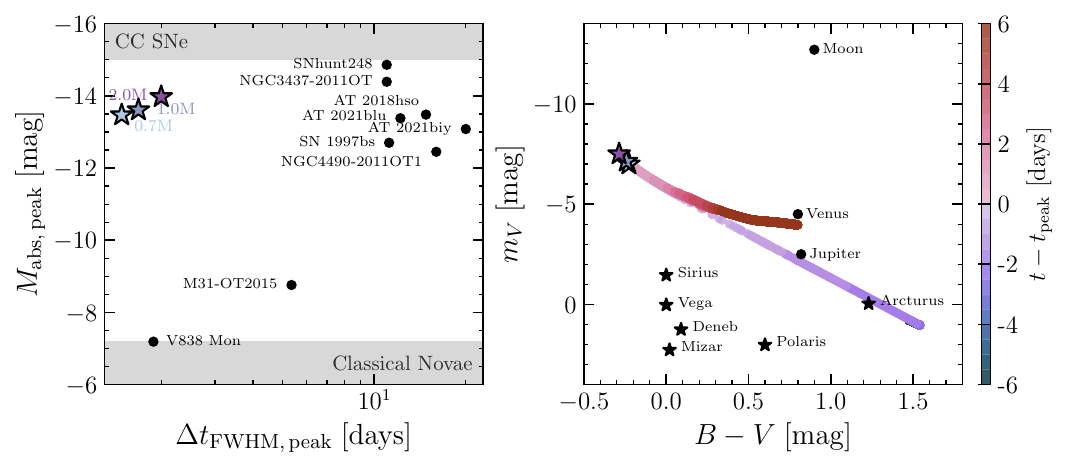}
    \caption{Left panel: Peak absolute magnitude versus the full width at half maximum (FWHM) duration of the blue peak for LRNe with well-sampled early-time observations \citep{Munari2002,Pastorello2019,Cai2022,Karambelkar2023}. Absolute magnitudes are reported in the Johnson--Cousins $V$ band, except for AT2018hso and AT2021blu, where ZTF-$g$ magnitudes from \citet{Karambelkar2023} are used. The colored stars denote the predicted shock-breakout and cooling transients from this work for companion masses of $0.7$, $1.0$, and $2.0\,M_\odot$. Right panel: Color--magnitude diagram showing the predicted evolution of the $1.0,M_\odot$ model compared with bright stars and familiar objects in the night sky. Black stars indicate the brightest stars, while black circles denote Solar System objects. Apparent magnitudes for the Moon, Venus, and Jupiter are taken from JPL Horizons \citep{Giorgini1996}, and stellar photometry is from \citet{Hoffleit1991}.}
    \label{fig:discussion}
    
\end{figure*}
\section{Discussion} \label{sec:Discussion}

\subsection{What Are We Going to See and When?}

Betelgeuse occupies a unique position among known and candidate LRN progenitors. Unlike previously observed LRNe, whose progenitor systems were either identified only after eruption or remained poorly constrained before outburst, Betelgeuse is nearby, extensively characterized, and observed prior to engulfment. Current observations constrain its radius, luminosity, evolutionary state, and likely binary configuration \citep{Goldberg2024,MacLeod2025,Howell2025}. Moreover, Betelgeuse is among the most evolved and extended progenitors currently associated with a future stellar merger \citep[e.g.,][]{Pastorello2019, Pastorello2023, Karambelkar2023, Matsumoto2022,Twum2026}.

Although LRNe are expected to occur in the Milky Way at a rate of approximately $0.1\,{\rm yr}^{-1}$ \citep{Kochanek2014,Howitt2020}, corresponding to roughly one event per decade, most occur at distances of several kiloparsecs and are discovered only after eruption. This rate is broadly consistent with the observationally inferred volumetric rate of $7.8^{+6.5}_{-3.7}\times10^{-5}\,{\rm Mpc}^{-3}\,{\rm yr}^{-1}$ \citep{Karambelkar2023}. The Betelgeuse-Siwarha system is at a distance of only $\approx 200$ pc. This system therefore represents a rare nearby realization of a phenomenon that has shaped stellar populations throughout the history of the Galaxy.

The extraordinary visibility of the event is illustrated in Figure~\ref{fig:discussion}. The right panel compares the predicted apparent brightness and color evolution of the transient to familiar objects in the night sky. The predicted breakout reaches apparent magnitudes of $m_V\approx-5.4$ to $-6.2$, making it several times brighter than Venus while remaining fainter than the full Moon. The transient would also substantially exceed the brightness of Sirius, the brightest star currently visible in the night sky. The figure also highlights the dramatic color evolution of the transient. Initially, the source remains relatively red while the shock propagates through the envelope. As shock breakout occurs, the transient rapidly becomes bluer than Sirius, reflecting the emergence of hot shock-heated material at the photosphere. The source then reddens again as the ejecta expands and cools during the subsequent cooling phase. Such pronounced changes in both luminosity and color would provide an unprecedented opportunity to monitor the earliest stages of a stellar merger in real time. 

It is important to note we do not account for extinction by circumstellar dust, although detailed models of Betelgeuse's dusty environment suggest that some attenuation may occur depending on the distribution and composition of the surrounding material \citep{Truong2022}. The predicted optical brightness and color evolution assume that the emerging radiation is not significantly obscured or reprocessed by circumstellar dust. Dust formed in material lost before engulfment could instead produce a substantially redder, optically fainter transient, potentially accompanied by significant infrared emission. We discuss this uncertainty further in Section \ref{sec:subsec}.

A natural question is when such an event might occur. In our binary-evolution calculations, the predicted time to engulfment ranges from approximately $4\times10^3$ to $7\times10^3$ yr for the considered companion masses, with the structure of the Betelgeuse progenitor at the onset of CEE remaining broadly similar. The primary effect of the companion mass here is that more massive companions inspiral in more rapidly because of stronger tidal coupling to Betelgeuse's extended envelope. The eventual engulfment therefore appears robust, although the precise timing still remains uncertain. In particular, the efficiency of tidal dissipation in the extended convective envelope of Betelgeuse remains poorly constrained. As emphasized by studies of stellar \citep{MacLeod2025} and planetary engulfment \citep{Yarza2023,Yarza2025}, modest changes in the tidal prescription can lead to substantial differences in the predicted merger timescale. Consequently, these times should therefore be viewed as order-of-magnitude estimates. Improved constraints on the companion orbit and tidal dissipation will ultimately determine when Betelgeuse enters CEE.

\subsection{How Does This Transient Compare to What We Have Seen?}
The early blue peak of LRNe is generally interpreted as cooling emission from shock-heated ejecta and is distinct from the longer-lived recombination-powered red peak (see \citealt{Kaminski2026} for an observational review). Here we compare our predicted Betelgeuse transients with the observed population, as seen in the left panel of Figure~\ref{fig:discussion}. The Betelgeuse models occupy the short-duration, luminous region of the observed blue-peak distribution. This is notable because Betelgeuse is among the most evolved and extended progenitors associated with a future LRN for which detailed pre-merger constraints are available \citep{Twum2026}. The predicted durations remain relatively short, spanning only $1.48$--$1.99$ days. Taken together, these results suggest that the diversity of LRNe blue peaks is shaped by more than the donor radius or the total available orbital energy. The luminosity and duration of the early emission also encode the depth of energy deposition, the mass and velocity of the shocked ejecta, and the coupling of the disrupted companion to the envelope \citep{Piro2013}. The early blue peak may therefore retain a direct memory of the companion and of the physics of its disruption.


Several caveats complicate a direct comparison with the observed LRN population. Many LRNe lack the high-cadence observations needed to fully characterize their earliest evolution \citep{Munari2002,Pastorello2019,Cai2022,Karambelkar2023}. Some systems may therefore have possessed brighter or shorter-lived blue peaks than were actually observed. Our calculations also neglect interaction with Betelgeuse's CSM. Such interaction is expected to broaden the breakout signal and reduce its peak luminosity, as seen in red-supergiant Type II supernovae \citep{Gonzalez2015, Ofek2014}. We also emphasize that our calculations are inherently one-dimensional and rely on various other assumptions, which are described in more detail in the following Section \ref{sec:subsec}. The durations reported here should therefore be regarded as lower limits.

Our ejecta masses also differ from expectations based on simple analytic common-envelope prescriptions. In the $0.7\,M_\odot$ case, the classical energy formalism predicts no $\alpha=1$ solution, yet the hydrodynamic simulation still ejects several solar masses of material. Conversely, for the $1.0$ and $2.0\,M_\odot$ cases, the analytic prescription predicts larger ejecta masses and substantially higher velocities than the simulations. These differences arise because the analytic formalism does not capture shock propagation, tidal disruption, or the mass loading of the ejecta by the disrupted companion. We discuss this comparison in detail in Appendix~\ref{app:alpha}. 

Taken together, these results suggest that the diversity of LRN blue peaks is shaped by more than donor radius or the total available orbital energy. The observed luminosity and duration also encode the depth of energy deposition, the structure of the engulfed companion, and the efficiency with which disruption energy and companion mass are coupled to the envelope. Betelgeuse therefore provides a particularly valuable test case for interpreting the early blue peaks of LRNe as diagnostics of CEE and stellar-merger physics.

The growing evidence that long secondary periods in red supergiants may arise from low-mass companions embedded in wide orbits raises the possibility that Betelgeuse is not unique \citep{Goldberg2024,Howell2025,MacLeod2025}. If this interpretation is correct, long-secondary-period red supergiants may represent an emerging population of future LRN progenitors. To assess how frequently systems similar to Betelgeuse are expected to undergo merger, we used the population-synthesis framework of \citet{Twum2026}, accounting for the observational uncertainties in the mass and radius of Betelgeuse as well as the inferred companion-mass range reported by \citet{MacLeod2025}. We find that Betelgeuse-like systems ($M=18.1_{-7.7}^{+6.8} \, M_\odot $, $R = 782 \pm 55 \,R_\odot$, $0.05 \leq q \leq 0.13$) comprise $9.8\%$ of merger progenitors assuming a flat mass-ratio distribution, $33.3\%$ under the binary parameter distributions of \citet{Sana2012}, and $8.0\%$ adopting the mass-dependent multiplicity prescriptions of \citet{MoeDiStefano2017}. This suggests that Betelgeuse-like mergers may represent a significant subclass of LRN progenitors rather than an exceptional evolutionary pathway.

\subsection{Comparison to Previous Work and Remaining Uncertainties \label{sec:subsec}}

Our calculations bridge several stages of stellar-merger evolution that are typically treated independently. Starting from self-consistent \mesa models of the Betelgeuse-Siwarha progenitor system, we follow the binary evolution to the onset of CEE, estimate the inspiral and tidal disruption of the companion, deposit both the orbital and disruption energy into the envelope, and compute the resulting shock propagation and breakout emission with radiation hydrodynamics. This framework directly links the properties of the pre-merger binary to the observable signatures of the transient.

Previous radiation-hydrodynamic studies have largely focused on the evolution of the ejecta following merger or mass ejection, whereas our calculations connect the ejecta properties directly to the pre-merger binary configuration and the disruption of the companion. For example,  \citet{Chen2024} and \citet{Kirilov2025} prescribe the initial ejecta properties and investigate the subsequent radiative evolution, demonstrating the importance of hydrogen recombination and the recession of the recombination front in shaping the optical light curve. In particular, \citet{Kirilov2025}  explore mergers both with and without a pre-existing CSM, finding that the contribution of recombination energy to the plateau luminosity and duration depends sensitively on the initial thermodynamic state of the ejecta  and its interaction with the CSM. These studies therefore complement our calculations by exploring the later radiative evolution of merger ejecta, while our work provides a physical connection between the binary interaction, energy deposition, and the properties of the emerging outflow. More recently, \citet{Hatfull2025} employed three-dimensional radiation hydrodynamics with companions represented by relaxed \mesa models to investigate merger-powered transients involving relatively compact stellar companions, demonstrating that companion structure can significantly influence the resulting light curve. In contrast, we focus on the engulfment and tidal disruption of low-mass companions within the extended envelope of a red supergiant and show how the deposition of both companion mass and disruption energy shapes the resulting shock-breakout transient. Our calculations therefore show that the shock-breakout and cooling emission are not determined solely by the donor star, but also retain a memory of the companion through the depth of disruption and the coupled deposition of mass and energy. This provides a direct observational link between the early transient and the physics of CEE.

Our calculations are designed primarily to follow the shock-breakout and early cooling emission and do not attempt to model the subsequent hydrogen-recombination plateau in detail. The ejecta masses and characteristic velocities obtained from our simulations nevertheless provide the quantities needed to estimate the expected plateau timescale using analytic recombination-wave models. A direct calculation of this phase would require additional microphysics not included in our present treatment. Moreover, the opacity changes dramatically as hydrogen recombines, from values characteristic of electron scattering in the ionized material to much smaller values in predominantly neutral gas.

Although the opacity tables used in our simulations account for changes in the ionization state of hydrogen, our current equation of state does not include the associated recombination energy in the thermodynamic evolution of the ejecta. Consequently, while the simulations capture the opacity changes associated with the recession of the recombination front, they do not self-consistently follow the release and radiative transport of hydrogen recombination energy, which can play an important role in regulating the luminosity and duration of the subsequent optical plateau. This limitation is not relevant for the early blue peak \citep{Antoni2025}, which is the primary focus of our calculations, but becomes important during the later recombination-dominated phase.

We estimate the luminosity and duration of the plateau using the numerical fits from \citet{Antoni2025}, for our simulated values of the ejecta mass $M_{\rm ej}$, ejecta kinetic energy $E_{\rm ej}$, and donor radius $R_*$. For companion masses of $0.7$, $1.0$, and $2.0\,M_\odot$, respectively, we obtain plateau luminosities of $L_{\rm plateau}=6.44\times10^{39}$, $7.51\times10^{39}$, and $1.22\times10^{40}\,{\rm erg\,s^{-1}}$, with corresponding durations of $t_{\rm plateau}=283$, $302$, and $306$ days. Furthermore, we can estimate the total energy that could be radiated in recombination photons using $E_{\rm ej} \sim (M_{\rm ej}/ m_p) *(13.6eV)* X_H$. Using the average hydrogen mass fraction of the ejecta $X_H \approx 0.66$, we obtain $E_{\rm ej} = 1.05 \times 10^{47}, 1.39 \times 10^{47}, \mathrm{\; and \;} 1.91 \times 10^{47}$ erg. Compared to the energy integrated over the plateau ($E _{\rm plateau} \sim L_{\rm plateau} \times t_{\rm plateau}$), which yields $1.57 \times 10^{47}, 1.96 \times 10^{47}, \mathrm{\; and \;} 3.23 \times 10^{47}$ erg, we find the ratio of total plateau energy to available recombination energy is $E_{\rm plateau} / E_{\rm ej} \approx 1.50, 1.41, \mathrm{\; and \;} 1.69$. Thus, hydrogen recombination could contribute a non-negligible fraction of emission to the plateau luminosity, so we do not extend our simulations to the plateau phase.

The present calculations nevertheless retain several important simplifications. Our first assumption is that the final stages of inspiral and tidal disruption occur rapidly compared to the dynamical response time of the envelope, allowing both the deposited mass and energy to be treated as impulsive \citep{2015ApJ...798L..19M,Everson2025}. In reality, the companion is expected to undergo an extended phase of orbital decay, mass transfer, and envelope restructuring before the final plunge \citep{MacLeod2015,MacLeod2017, Clayton2017, MacLeod2018a, MacLeod2018b, Fragos2019}. Allowing the envelope to respond dynamically during the inspiral should modify the distribution of CSM around the donor star therefore also the subsequent shock propagation and luminosity \citep{Metzger2017}. Having established these calculations as a baseline, we intend to study those effects in future work. 

Our models also neglect the interaction of the ejecta with Betelgeuse's CSM. The star has experienced substantial mass loss throughout its evolution as a red supergiant \citep{Yoon2010}, producing a dusty and structured CSM that is likely to modify the emergent transient. As demonstrated in radiation-hydrodynamic models of Type II supernovae, interaction with a dense CSM broadens the light curve, reduces the peak luminosity, and can substantially alter the color evolution (\citealt{Suzuki2019}, \citealt{Dessart2026}; see also \citealt{Dessart2024} for a review). These changes in the observational signatures are also seen in models of failed supernovae with red supergiant progenitors \citep{Tsuna2025}.  Consequently, the peak luminosities reported here should be regarded as approximate upper limits, while the breakout durations likely represent lower limits.

If sufficient dust is present in the pre-outburst CSM, absorption and scattering could substantially suppress the optical and ultraviolet emission predicted by our calculations, while reprocessing the absorbed radiation into the infrared. Consequently, the intrinsically luminous blue peak may appear considerably redder and fainter at optical wavelengths, even if the total radiated energy is largely conserved. \citet{MacLeod2022} estimated the pre-outburst dust optical depth of merging stellar systems. Under their assumptions the predicted V-band reddening is $A_V \lesssim 0.16$~mag  (greatest for the largest companion masses, see their equations 19 and 20) implying a slightly reddened, but not obscured transient. However, the magnitude of this effect depends on the mass, geometry, and optical depth of the circumstellar material, as well as how much the initial radiation pulse destroys nearby dust. Because our calculations do not include circumstellar dust formation, destruction, or radiative reprocessing, we are therefore predicting the intrinsic emission rather than its wavelength-dependent appearance in a potentially-dusty environment.

A second limitation is the assumption of spherical symmetry. The merger with an RSG is inherently three-dimensional, involving asymmetric energy deposition, convection, large-scale flows, and anisotropic ejecta \citep{Ma2024, Ma2025, Jiang2023, Nibbs2025, Lau2025, Hutchinson-Smith2026,Schreier2026}. Multidimensional simulations of shock breakout from red supergiants further demonstrate that surface convection broadens the emerging radiation pulse and extends its duration relative to one-dimensional calculations \citep{Goldberg2022}. Our predicted breakout durations should therefore be regarded as representative of the underlying diffusion timescale rather than precise observational predictions.

An alternative interpretation of Betelgeuse's rapid rotation and surface abundance anomalies is that these are signatures of a previous merger, rather than the presence of a binary companion \citep{Wheeler2017,Chatzopoulos2020,Shiber2024}. The growing evidence for a presently orbiting companion complicates this picture: if both interpretations are correct, Betelgeuse would require a more complex interaction history, potentially involving a hierarchical multiple system. We do not explore such scenarios here and instead adopt the presently inferred Betelgeuse–Siwarha binary as our initial condition.

Despite these simplifications, we expect the principal trends identified here to remain robust. In particular, the dependence of the breakout luminosity and duration on companion mass, disruption radius, and the coupled deposition of mass and energy reflects basic energetics that should persist in more realistic multidimensional calculations. Future work should therefore combine self-consistent inspiral calculations with multidimensional radiation hydrodynamics and realistic models of Betelgeuse's CSM to quantify how these effects modify the observable transient. The qualitative dependence on companion mass and disruption radius found here is nevertheless expected to remain robust, making the early breakout emission a powerful probe of CEE in massive stars.

\subsection{Implications for the Future Evolution of Betelgeuse}

The consequences of companion engulfment extend well beyond the LRN itself. Our simulations eject $6$--$11\,M_\odot$ of hydrogen-rich material, removing a substantial fraction of Betelgeuse's envelope while simultaneously depositing several solar masses into the surrounding CSM. Although the merger does not completely strip the hydrogen envelope, it is expected to significantly modify both the structure of the progenitor and the environment into which the eventual supernova will explode.

Immediately following the merger, the remnant will undergo substantial thermal and structural readjustment as orbital energy is redistributed and part of the hydrogen envelope is removed \citep{Schneider2025, Patton2025, Picco2026}. The eventual post-merger state of Betelgeuse will depend not only on the amount of mass removed, but it may also depend on the degree of mixing induced during the merger \citep{Ivanova2002, Gaburov2008, Menon2024}. For all three models, the disruption radius lies interior to the base of the convective hydrogen envelope, $r_{\rm dis}<r_{\rm H,conv}$, implying that companion material may be deposited within radiative layers deep in the stellar interior. If hydrogen-rich material is mixed into sufficiently hot regions, the resulting modification of the composition gradient could influence subsequent shell burning and potentially trigger additional nuclear burning. Following this compositional and thermal readjustment requires post-merger stellar-evolution calculations and lies beyond the scope of the present work.

The removal of several solar masses of envelope material will influence the eventual core-collapse supernova. The mass and structure of the hydrogen envelope play a central role in determining the diffusion timescale, ejecta mass, photospheric velocities, and light-curve evolution of core-collapse supernovae \citep{Podsiadlowski1992,Eldridge2008,Sana2012,Smith2014,Taddia2018,Lu2025,Yadavalli2026, Ercolino2026}. In our simulations, $6.15$--$11.19\,M_\odot$ of material is ejected with characteristic velocities of approximately $194$--$217\,{\rm km\,s^{-1}}$. The material remains relatively close to the star on stellar evolutionary timescales, creating a CSM that differs markedly from that produced by steady red-supergiant winds. To find the potential merger-generated CSM location when a supernova could occur, we can estimate the location of the ejecta at $t = 10^4$--$10^5$ years. If we look at the fiducial $1.0 M_\odot$ model, we will find $R_{\rm CSM} \sim v_{\rm ej} t \approx 6.5 \times 10^{18}$--$10^{19}$ cm ($\approx 2$--$20 \; \mathrm{pc}$). A supernova blast wave with $v_{SN} = 2000 \; \mathrm{km \,s}^{-1}$ will then take $t_\mathrm{sweep} \approx R_\mathrm{CSM} / v_{SN} \approx 100$--$1000$ years to sweep up this shell. The eventual supernova may therefore explode into a medium shaped by the merger itself. Interaction between the supernova ejecta and this merger-generated CSM could alter the luminosity, color evolution, and spectroscopic appearance of the explosion \citep{Chevalier1994,Moriya2011,Smith2017}. Consequently, the eventual supernova may differ significantly from that expected for an isolated Betelgeuse, even if the star ultimately retains sufficient hydrogen to remain spectroscopically Type II.

More broadly, these results suggest that stellar mergers may influence core-collapse supernovae through two complementary channels. First, they modify the internal structure of the progenitor through the removal and redistribution of envelope material. Second, they shape the CSM into which the supernova ultimately explodes. Betelgeuse therefore offers an opportunity to investigate how a stellar merger can influence not only the transient it immediately produces, but also the final stages of massive-star evolution leading to core collapse. More broadly, Betelgeuse connects the physics of binary interaction, CEE, stellar mergers, and core-collapse supernovae within a single nearby system. The calculations presented here establish the first predictive framework linking the future merger of Betelgeuse to its observable transient signatures and their implications for the star's final evolution.

\begin{acknowledgments}
We gratefully acknowledge helpful discussions with M. Cantiello, A. Dupree, R. Everson, R. Foley, J. Goldberg, T. Hutchinson-Smith, Y.-F. Jiang, A. Rosselli-Calderon, R. Yarza, and A. Manzo. We acknowledge use of the lux supercomputer at UC Santa Cruz, funded by NSF MRI grant AST 1828315.  This research was made possible through funding from the Lamat Institute \citep{2025NatAs...9.1770Q} and UC Santa Cruz through the Heising-Simons Foundation, and NSF grants AST-1852393, AST-1911206, AST-2150255, AST-2206243 and 2447606. The Flatiron Institute is supported by the Simons Foundation.

\software{astropy \citep{2013A&A...558A..33A,2018AJ....156..123A,2022ApJ...935..167A},  
          Cloudy \citep{2013RMxAA..49..137F}, 
          Source Extractor \citep{1996A&AS..117..393B}
          }
\end{acknowledgments}

\appendix


\section{Comparison with the Classical $\alpha$-Formalism \label{app:alpha}}

\begin{figure}[ht]
\centering
\includegraphics[width=4.5in]{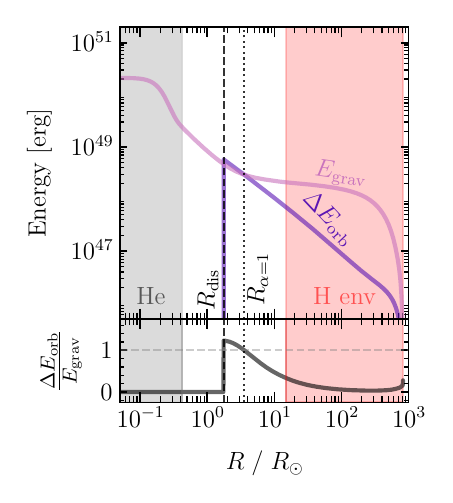}                  
\caption{Comparison of the cumulative orbital energy released during inspiral and the binding energy of the Betelgeuse envelope. Top panel: orbital energy release, $\Delta E_{\rm orb}$ (purple), and envelope binding energy, $E_{\rm grav}$ (pink), as functions of radius for the fiducial $1.0M_\odot$ companion. The dotted vertical line marks the classical $\alpha=1$ solution, where $\Delta E_{\rm orb}=E_{\rm grav}$, while the dashed vertical line indicates the tidal disruption radius of the companion. Bottom panel: same as the top panel, but in units of $\Delta E_{\rm orb}/E_{\rm grav}$. The figure provides the energetic framework underlying the classical $\alpha$-formalism and serves as a reference for comparison with the hydrodynamic calculations.}
\label{fig:dEorb_Egrav}
\end{figure}

To place our hydrodynamic calculations in the context of traditional common-envelope prescriptions, we compare the simulated ejecta masses and velocities to those predicted by the standard $\alpha$-formalism. Following \citet{Twum2026}, the orbital energy released during inspiral to a radius $r$ is
\begin{equation}
\Delta E_{\rm orb}(r)
=
\frac{G M_\ast m_c}{2R_\ast}
-
\frac{G M_{\rm enc}(r)m_c}{2r},
\end{equation}
where $M_\ast$ and $R_\ast$ are the mass and radius of Betelgeuse at the onset of engulfment, $m_c$ is the companion mass, and $M_{\rm enc}(r)$ is the enclosed mass of Betelgeuse at radius $r$. The envelope binding energy exterior to radius $r$ is denoted by $E_{\rm grav}(r)$. Within the classical energy formalism, envelope ejection becomes possible when $\alpha \equiv \Delta E_{\rm orb}/E_{\rm grav} \gtrsim 1$. The radius at which this condition is first satisfied defines the $\alpha=1$ solution. Material exterior to this location is assumed to become unbound. 

The characteristic ejecta velocity predicted by the $\alpha$-formalism is estimated by assuming that the released orbital energy is distributed uniformly among all material exterior to the $\alpha=1$ radius. Equating the available orbital energy to the kinetic energy of the ejecta yields
\begin{equation}
v_{\rm ej,\alpha}
=
\left(
\frac{2\Delta E_{\rm orb}}
{M_{\rm ej,\alpha}}
\right)^{1/2},
\label{eq:v_ej}
\end{equation}
where $M_{\rm ej,\alpha}$ is the mass exterior to the $\alpha=1$ coordinate. This estimate should be interpreted as a characteristic velocity associated with the available specific energy rather than a prediction of the full ejecta velocity distribution.

Figure~\ref{fig:dEorb_Egrav} illustrates the standard $\alpha$-formalism applied to the Betelgeuse progenitor at the onset of engulfment. The figure compares the cumulative orbital energy released during inspiral with the binding energy of the overlying envelope as a function of radius for each companion mass. The locations where these curves intersect define the classical $\alpha=1$ solutions. This construction enables an analytic estimate of the mass available for ejection and the characteristic ejecta velocity expected from the energy formalism. These predictions serve as a useful reference point for interpreting the hydrodynamic results presented in Section~\ref{sec:Results} and should be contrasted with the physical energy-deposition profiles shown in Figure~\ref{fig:energy_deposition}, where both tidal-disruption energy and companion mass loading are explicitly included.

For the $1.0M_\odot$ companion, the $\alpha=1$ condition is reached at $3.70R_\odot$, corresponding to an ejecta mass of $12.67M_\odot$ and a characteristic velocity of $759~{\rm km~s^{-1}}$. For the $2.0M_\odot$ companion, the crossing occurs at $9.44R_\odot$, yielding an ejecta mass of $12.51M_\odot$ and a characteristic velocity of $483~{\rm km~s^{-1}}$. In contrast, no $\alpha=1$ solution exists for the $0.7M_\odot$ companion because the released orbital energy never exceeds the binding energy of the overlying envelope. 

The simulations, for comparison, yield unbound masses of $6.15M_\odot$, $8.11M_\odot$, and $11.19 \; M_\odot$ for the $0.7M_\odot$, $1.0M_\odot$, and $2.0M_\odot$ companions, respectively. The corresponding mass-weighted ejecta velocities are $194$, $207$, and $217~{\rm km~s^{-1}}$. The discrepancy is most striking for the $0.7M_\odot$ companion. Although no $\alpha=1$ solution exists, the hydrodynamic simulation still ejects approximately $6.15M_\odot$ of material with a mass-weighted velocity of $194~{\rm km~s^{-1}}$. The absence of an $\alpha=1$ solution therefore does not imply the absence of mass ejection. Instead, the simulations demonstrate that shocks launched during inspiral and tidal disruption can transfer energy and momentum to the outer envelope in a manner that is not captured by global energy-balance arguments.

The origin of these differences is twofold. First, in the $\alpha$-formalism, the companion contributes orbital energy to the envelope but does not add mass to the ejecta. In our simulations, by contrast, the companion is tidally disrupted within the envelope, depositing both energy and mass into the surrounding gas. Although disruption increases the total energy available to drive the outflow, the accompanying mass loading distributes this energy over a larger amount of material, reducing the specific energy relative to a calculation with the same energy deposition but no added mass. Second, the propagation of the shock through the stratified stellar envelope produces a nonuniform distribution of energy and velocity among the ejecta. The velocity estimate derived from the $\alpha$-formalism in Eq.~\ref{eq:v_ej}, by contrast, effectively distributes the available kinetic energy uniformly per unit mass. Our radiation-hydrodynamic calculations therefore capture both the effects of companion mass deposition and the redistribution of energy by shock propagation, which can lead to different characteristic ejecta velocities than those inferred from a global energy-budget argument. We emphasize that both approaches remain simplified descriptions of an intrinsically multidimensional common-envelope interaction.

Three-dimensional simulations show that orbital energy is deposited through spiral shocks and gravitational wakes and is not shared instantaneously throughout the envelope. In many cases, this delayed and anisotropic energy deposition leads to greater envelope ejection than predicted by simple energy-balance arguments. For example, \citet{Tenley2024} found that an embedded object, modeled as a point mass, can continue depositing orbital energy and driving shocks below the nominal $\alpha=1$ radius, leading to substantially more mass ejection than predicted by classical energy-balance prescriptions. Similar behavior has been found in other multidimensional studies of CEE, where shock propagation and anisotropic energy transport play a central role in determining the final ejecta properties \citep[e.g.,][]{2023LRCA....9....2R,2026arXiv260527579H}.

Our models differ from these calculations in that they explicitly include the tidal disruption and thermalization of a stellar companion but treat the subsequent evolution in spherical symmetry. Consequently, the $\alpha$-formalism likely overestimates the specific energy and characteristic velocity of the ejecta by neglecting the disrupted companion mass, whereas our one-dimensional treatment cannot capture the anisotropic and time-dependent redistribution of energy and momentum during inspiral. Existing three-dimensional common-envelope simulations generally do not include the tidal disruption and mass loading considered here. Fully three-dimensional radiation-hydrodynamic calculations that incorporate companion disruption will therefore be required to determine the ultimate ejecta masses, velocity distributions, and observational signatures of Betelgeuse-like mergers.

\section{Additional Simulation Figures}
Here, we present figures showing the shock propagation through time for the other simulated companion masses, specifically $0.7 \,M_\odot$ and $2.0 \,M_\odot$, similar to Figure \ref{fig:time_evolve} for the fiducial $1.0 \,M_\odot$ model.

\begin{figure}[ht]
    \centering
    \includegraphics[width=7.0in]{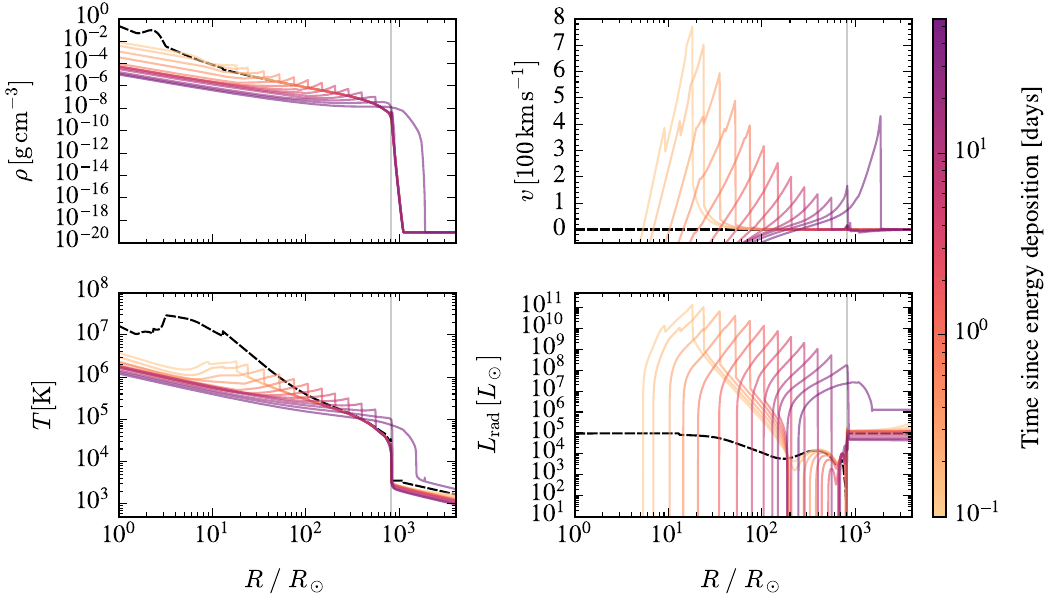}
    \caption{Evolution of the $0.7\,M_\odot$ companion model following the deposition of $E_{\rm total}=6.29\times10^{48}$ erg.
}
    \label{fig:time_evolve_0.7}
\end{figure}

\begin{figure}[ht]
    \centering
    \includegraphics[width=7.0in]{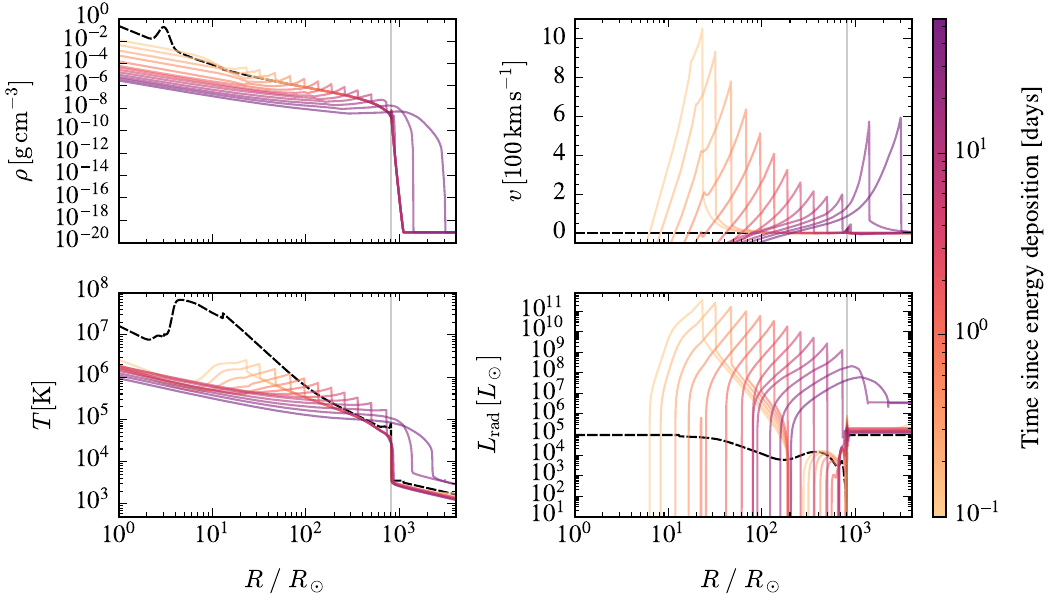}
    \caption{Evolution of the $2.0 \,M_\odot$ companion model following the deposition of $E_{\rm total}=1.39\times10^{49}$ erg.
}
    \label{fig:time_evolve_2.0}
\end{figure}

\pagebreak

\bibliography{sample701}{}
\bibliographystyle{aasjournalv7}



\end{document}